\documentclass[paper]{agujournal2019}
\usepackage{url}
\usepackage{lineno}
\usepackage[inline]{trackchanges} 
\usepackage{soul}
\usepackage{color}
\usepackage{amsmath}
\usepackage{amssymb}

\draftfalse

\newcommand{\ssr}{   {Space Sci. Rev. }}

\newcommand{\jgr}{   {J. Geophys. Res.}}
\newcommand{\grl}{   {Geophys. Res. Lett.}}
\newcommand{\apj}{   {Astrophys. J.}}
\newcommand{\apjl}{   {Astrophys. J. Lett.}}
\newcommand{\prl}{   {Phys. Rev. Lett.}}

\newcommand{\araa}{   {Annual Review of Astronomy and Astrophysics}}

\graphicspath{{./}{figures/}}

\journalname{Geophysical Research Letters}

\begin{document}


\title{Relativistic and Ultra-Relativistic Electron Bursts in Earth's Magnetotail Observed by Low-Altitude Satellites}

\authors{Xiao-Jia Zhang  \affil{1,2}, Anton V. Artemyev \affil{2}, Xinlin Li\affil{3,4}, Harry Arnold\affil{5}, Vassilis Angelopoulos \affil{2}, Drew L. Turner\affil{5}, Mykhaylo Shumko \affil{5}, Andrei Runov \affil{2}, Yang Mei\affil{3,4}, Zheng Xiang\affil{4}}
\affiliation{1}{Department of Physics, The University of Texas at Dallas, Richardson, TX, USA}
\affiliation{2}{University of California, Los Angeles, Los Angeles, USA}
\affiliation{3}{Laboratory for Atmospheric and Space Physics, University of Colorado Boulder, Boulder, CO, USA}
\affiliation{4}{Department of Aerospace Engineering Sciences, University of Colorado Boulder, Boulder, CO, USA}
\affiliation{5}{The Johns Hopkins University Applied Physics Laboratory, Laurel, MD USA}

\correspondingauthor{Xiao-Jia Zhang}{xjzhang@utdallas.edu}

\begin{keypoints}
\item We report observations of relativistic and ultra-relativistic electrons in near-Earth magnetotail
\item We show energy spectra of relativistic electron bursts in the magnetotail
\item We discuss potential mechanisms responsible for the formation of relativistic and ultra-relativistic electrons
\end{keypoints}

\begin{abstract}
Earth's magnetotail, a night-side region characterized by stretched magnetic field lines and strong plasma currents, is the primary site for the release of magnetic field energy and its transformation into plasma heating and kinetic energy plus charged particle acceleration during magnetic reconnection. In this study, we demonstrate that the efficiency of this acceleration can be sufficiently high to produce populations of relativistic and ultra-relativistic electrons, with energies up to several MeV, which exceeds all previous theoretical and simulation estimates. Using data from the low altitude ELFIN and CIRBE CubeSats, we show multiple events of relativistic electron bursts within the magnetotail, far poleward of the outer radiation belt. These bursts are characterized by power-law energy spectra and can be detected during even moderate substorms. 
\end{abstract}

\section*{Plain Language Summary}
Charged particle acceleration during magnetic field line reconnection is a universal process occurring in various space plasma environments. Traditionally, theoretical and simulation models of this acceleration are verified using data from the reconnection region in the near-Earth magnetosphere, where in-situ spacecraft measurements are most accessible. In this study, we demonstrate that the efficiency of this acceleration can  significantly exceed previous estimates, leading to the formation of electron populations with energies up to several MeV, even in regions where thermal electron energies are below 1 keV. These observations of highly energetic electron bursts are made available by new low-altitude CubeSat missions monitoring magnetotail electron fluxes.

\section{Introduction}\label{sec:intro}
The primary mechanism of charged particle acceleration in Earth's magnetotail, the night-side region of the magnetosphere, is presumably related to near-Earth magnetotail reconnection \cite{Baker96,Angelopoulos08,Paschmann13}. Numerous theoretical and observational investigations have focused on charged particle acceleration in near-Earth reconnection, because it determines the energy flux transported into the inner magnetosphere, which further supports the ring current and outer radiation belt \cite<see discussions in>[]{Angelopoulos20,Lin21:ANGIE3D_CIMI,Jun21:emic,Hua23,Kim23}. 

Numerical simulations of electron acceleration in realistic magnetotail magnetic field configurations \cite<see>[]{Birn04:pop, Maha11NatPh, Sorathia18} usually combine the test particle approach (due to the negligible feedback of the high-energy (non-thermal), low-flux population on electromagnetic fields) with global MHD models that can reproduce the main elements of magnetotail dynamics. These simulations demonstrate the dominant role of adiabatic (Fermi and betatron) mechanisms in electron energization, which can accelerate electrons up to $100-200$keV in the magnetotail \cite{Zhou11:lsk,Birn13:jgr}. Consistent with these model results, spacecraft observations confirm that effective electron acceleration due to magnetotail reconnection is largely contributed by electron acceleration at dipolarization fronts \cite{Vaivads11,Fu13:NatPh}, which are the leading edges of fast plasma flows carrying enhanced magnetic fields and strong electric fields \cite{Nakamura02, Runov09grl, Sitnov09}. These fronts can heat pre-accelerated electrons adiabatically via magnetic field line compression \cite{Maha11NatPh,Maha13:lsk,Pan12}, producing electrons with energies $\sim 100-300$keV \cite{Fu19:apjl_rx}. Although further electron acceleration up to $\sim 1$ MeV is possible during plasma injections into the inner magnetosphere \cite{Sorathia18,Turner21:grl_mms&rbsp}, existing magnetotail models mostly focus on reproducing sub-relativistic ($100-200$ keV) electron fluxes. This upper limit at sub-relativistic energies in magnetotail simulations is likely dictated by available observations: majority of energetic electron measurements in the outer magnetosphere are limited to sub-relativistic energies, $\sim 200-300$keV \cite{Imada07,Oieroset02,Retino08,Huang12:grl_rx,Cohen21,Ergun20:observations}. But in reality, can electron acceleration in the magnetotail be more effective, and what are the actual upper energies of accelerated electrons? Historical observations in the distant tail by ISEE-3 \cite{Slavin92,Richardson93} and in the middle tail by Imp 8 \cite{Baker&Stone77} reveal the presence of bursts of relativistic electrons, with energies of $\sim 0.2-2$MeV, although the geomagnetic context and origin of these bursts remain largely unknown. 

This question is challenging to address with near-equatorial observations, because of the quite transient nature of electron acceleration in the magnetotail and the typical upper energy limit of energetic particle detectors onboard magnetotail missions  \cite{Wilken01,Angelopoulos08:sst,Blake16}, which lack high resolution in relativistic energies. A promising alternative is the new generation of low-altitude CubeSat missions with high energy resolutions. These CubeSats can quickly cross the field lines conjugate to the entire near-Earth's magnetotail within a few minutes, effectively monitoring energetic electron populations in this region \cite{Artemyev22:jgr:ELFIN&THEMIS}. Here we analyze several events where relativistic (above $500$keV) and ultra-relativistic (several MeV) electron bursts were observed in Earth's magnetotail by Colorado Inner Radiation Belt Experiment (CIRBE) \cite{Li24:grl:CIRBE} and Electron Losses and Fields Investigation (ELFIN) \cite{Angelopoulos20:elfin} CubeSats. These observations likely represent the first clear separation of electron populations accelerated in the magnetotail and relativistic electron leakage from the outer radiation belt. We analyze the electron energy spectra and discuss possible mechanisms for such effective electron acceleration. 

\section{ELFIN observations}\label{sec:elfin}
\begin{figure}[!htbp]
    \centering
    \includegraphics[width=1.0\linewidth]{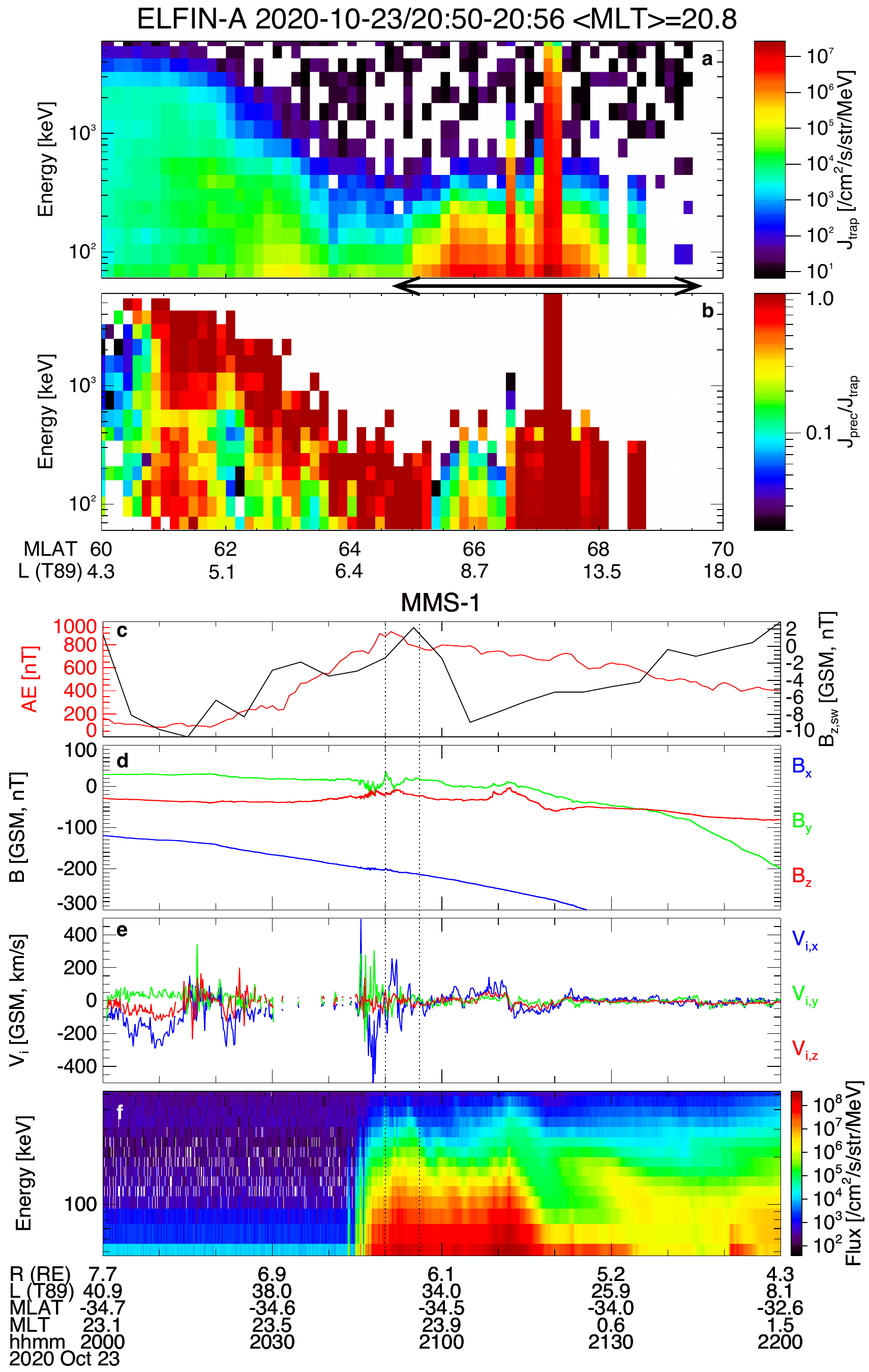}
    \caption{Overview of one ELFIN event with relativistic electron observations in near-Earth magnetotail on 23 October 2020, at $\sim$ 20:53 UT.
    Panels (a, b) show locally trapped electron fluxes and precipitating-to-trapped flux ratio (versus magnetic latitude). The black arrow marks the plasma sheet region along the ELFIN track. Panel (c) shows AE index and solar wind $B_z$ for the 2h interval embedding the ELFIN event. Panels (d,e,f) show MMS observations in the magnetotail: GSM magnetic field from \cite{Russell16:mms}, GSM plasma flows from \cite{Young16:mms}, and energetic electron spectra from \cite{Blake16}. }
    \label{fig1}
\end{figure}

The two identical ELFIN CubeSats are equipped with an energetic particle detector (EPD) that measures electrons within the $[50,6000]$keV energy range across sixteen logarithmically distributed channels, with full pitch-angle coverage ($22.5^\circ$ angular bin) and 3s spin resolution \cite{Angelopoulos20:elfin}. The pitch-angle resolution allows for the separation of locally trapped fluxes ($j_{trap}$) from precipitating (within the bounce loss cone) fluxes ($j_{prec}$), where $j_{trap}$ is outside of the bounce loss cone and $j_{prec}$ is within it. The ratio $j_{prec}/j_{trap}$ is useful in determining the specific magnetospheric region along the ELFIN track: the outer radiation belt is characterized by high $j_{trap}$ of relativistic electrons and a low $j_{prec}/j_{trap}$ ratio, with transient bursts of $j_{prec}/j_{trap}$ due to wave driven electron precipitation; in contrast, the plasma sheet region exhibits significant $j_{trap}$ only below $\sim 200-300$keV and a $j_{prec}/j_{trap}\sim 1$ due to strong magnetic field line curvature scattering \cite<see details in>{Mourenas21:jgr:ELFIN,Angelopoulos23:ssr}. During its routine operation from December 2019 to September 2022, ELFIN detected several events with bursty increases of significant $j_{trap}$ at energies up to a few MeVs within the plasma sheet. One such event is shown in Fig. \ref{fig1}(a,b): the black arrow marks the nominal plasma sheet region with electron fluxes at moderate energies ($<300$keV). Within this region, there is a burst of relativistic fluxes with energies reaching up to $\sim5$MeV. This burst is well separated from the outer radiation belt region and cannot be associated with the relativistic electron leakage (outward diffusion); instead, it is more likely generated by a local and highly effective acceleration mechanism. 
The inner edge of the plasma sheet, the electron isotropy boundary, as characterized by energy-latitude dispersion of $j_{prec}/j_{trap}\sim 1$ \cite<see>{Sergeev12:IB,Wilkins23}, is located within $[60,64^\circ]$, well equatorward from the relativistic electron burst. During this burst, the electron fluxes exceed the upper limit of ELFIN EPD, leading to a high background level \cite<see>[for discussion of this effect during relativistic microbursts measured by ELFIN in the outer radiation belt]{Zhang22:microbursts}. For spins with such a high background level, $j_{prec}/j_{trap}$ can show misleading variations, so we set it to $1$.

The ELFIN relativistic burst event in Fig. \ref{fig1}(a,b) was observed when MMS \cite{Burch16} was in the night-side magnetosphere, detecting substorm electron injections as evidenced in energetic electron fluxes (note that MMS was 3 hours of MLT closer to midnight than ELFIN A): panels (d-f) show magnetic field perturbations \cite{Russell16:mms}, plasma flow speed \cite{Young16:mms}, and energetic electron fluxes \cite{Blake16} typical for substorm injections observed at latitudes away from the equator \cite<see, e.g., discussion in>{Gabrielse19}. Therefore, ELFIN observations are associated with substorm activities (as confirmed by the AE index in panel (c)), including magnetic field-line reconnection, charged particle acceleration, and subsequent injection into the inner magnetosphere \cite<see>[and references therein]{Birn12:SSR,book:Gonzalez&Parker,Sitnov19}. However, MMS' energetic electron detector does not capture an increase in relativistic electron fluxes (with the main flux increase being below $500$keV), suggesting that the ELFIN-detected electron burst is likely spatially localized in MLT or may not reach lower $L$-shells (although MMS is projected far in the magnetotail due to a large $B_x$ magnitude, the significant uncertainties of the magnetic field model in the night-side injection region suggest that it might be mapped closer to its instantaneous location at $\sim 6.7R_E$), possibly being scattered and lost into the atmosphere. 

\begin{figure}[!htbp]
    \centering
   \includegraphics[width=1.0\linewidth]{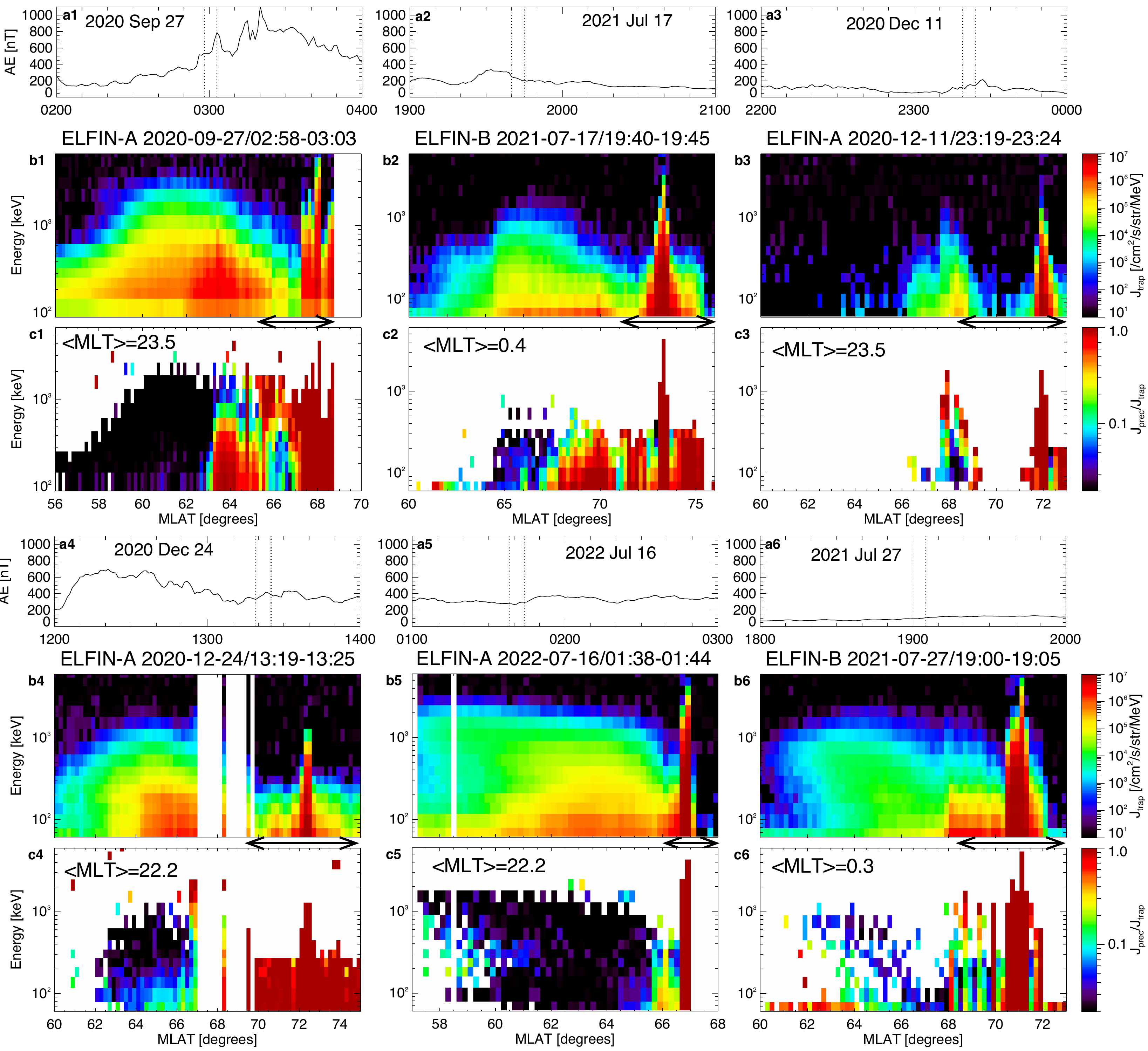}
    \caption{Overview of six ELFIN events with relativistic electron observations in near-Earth magnetotail: on 27 September 2020 at 03:02UT, 17 July 2021 at 19:45 UT,
    11 December 2020 at 23:20UT, 24 December 2020 at 13:23UT, 16 July 2021 at 19:05UT, and 27 July 2021 at 01:42UT. For each event, we show (a) the AE index during the 2h interval embedding ELFIN observations, and ELFIN measurements of (b) locally trapped fluxes and (c) precipitating-to-trapped flux ratio. Black arrows mark the plasma sheet region during each orbit.}
    \label{fig2}
\end{figure}

Figure \ref{fig2} shows six examples of similar events from ELFIN, where bursts of relativistic electron fluxes up to $5$MeV were detected well within the magnetotail, poleward from the outer radiation belt (black arrows in each event mark the latitudinal range of the magnetotail, poleward from the outer radiation belt). These events are associated with substorm activities (see AE profiles), but many of them occur during moderate AE levels ($< \sim 300$ nT). Therefore, we can infer that electron acceleration to relativistic energies in the magnetotail does not necessarily require specific geomagnetic conditions (e.g., intense substorm activity). However, due to limited ELFIN telemetry \cite<a few orbits per day; see details in >{Tsai24:review}, estimating the occurrence rate of such relativistic bursts is challenging in the ELFIN dataset. Moreover, all detected relativistic bursts are associated with very high background levels of the ELFIN EPD \cite<see discussion in>{Angelopoulos23:ssr, Tsai24:review}. Thus, a detailed analysis of the electron spectra during such events cannot be conducted using the ELFIN dataset. However, it reveals an important characteristic of these bursts: their isotropic pitch-angle distributions, with $j_{prec}/j_{trap}\sim 1$. Therefore, measurements of locally trapped fluxes are sufficient to (1) estimate the precipitating relativistic electron fluxes from the magnetotail, and (2) compare with equatorial flux measurements from missions like MMS and THEMIS. Such isotropy is expected because electrons at these high energies are rapidly scattered by magnetic field line curvature in the magnetotail current sheet.

\section{CIRBE observations}\label{sec:cirbe}
CIRBE is a 3U CubeSat equipped with the Relativistic Electron and Proton Telescope integrated little experiment-2 (REPTile-2) \cite{khoo2022challenges,Li22:cirbe},  measuring electrons within the $[250,6000]$keV energy range across sixty channels with a time resolution of 1 sec \cite{Li24:grl:CIRBE, Li24:CIRBE&AGU}. REPTile-2 has a field-of-view of 51$^{\circ}$ and a look direction nearly perpendicular to the background magnetic field, making it well-suited to measure perpendicular (90$^{\circ}$) fluxes, i.e., the locally-trapped flux, $j_{trap}$ \cite<see discussion in>{zhang2020long}. We analyzed CIRBE measurements during the prolonged storm interval, 4-7 November 2023, when CIRBE detected multiple relativistic electron bursts in near-Earth magnetotail. Although CIRBE does not resolve pitch-angle distributions, we may use the analogy of CIRBE and ELFIN measurements to distinguish the magnetotail region. To confirm that CIRBE's measurements of relativistic bursts are within the magnetotail, we also compare these measurements with precipitating and trapped electron fluxes measured by one of nearby POES satellites \cite{Evans&Greer04}, which are traditionally used to localize the magnetotail/outer radiation belt transition region \cite<e.g.>{Sergeev12:IB}. Two main advantages of CIRBE measurements are (1) excellent telemetry with data available from almost each orbit ($\sim 90$min), (2) high energy resolution and sensitivity, allowing for a reliable measurement of the spectrum of relativistic electron bursts. 

\begin{figure}[!htbp]
    \centering
    \includegraphics[width=1.0\linewidth]{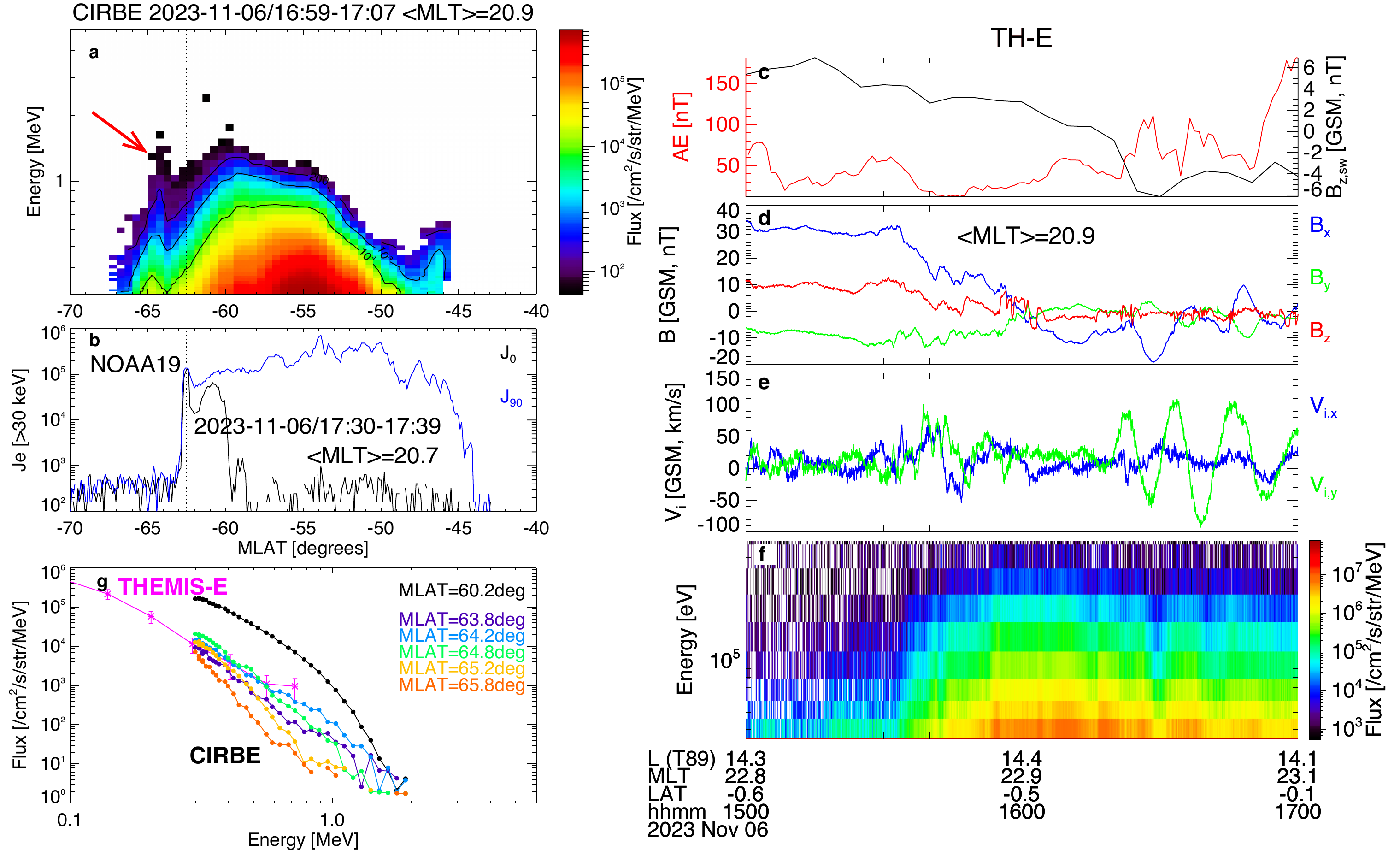}
    \caption{Overview of one CIRBE event with relativistic electron observations in near-Earth magnetotail on 06 November 2023, at $\sim$17:05 UT. Panel (a) shows locally trapped electron fluxes (versus magnetic latitude). Panel (b) shows POES observations of trapped and precipitating fluxes of $>30$keV electrons (vertical dashed line shows electron isotropy boundary). Panel (c) shows the AE index and solar wind $B_z$ for the 2h interval embedding the CIRBE event. Panels (d,e,f) show THEMIS observations in the magnetotail: GSM magnetic field from \cite{Auster08:THEMIS}, GSM plasma flows from \cite{McFadden08:THEMIS}, and energetic electron spectra from \cite{Angelopoulos08:sst}. Panel (g) shows the 1D electron spectra measured by CIRBE during relativistic electron bursts in the magnetotail, in comparison with THEMIS measurements during the substorm injection. }
    \label{fig3}
\end{figure}

Figure \ref{fig3}(a) shows CIRBE observations of an relativistic burst event: the peak of relativistic electron fluxes ($\sim64.5^\circ$ magnetic latitude) is well separated from the outer radiation belt region (see the black contour lines). 
Note that in the absence of measurements of $<250$keV electron fluxes \cite<i.e., the dominant energetic electron population in the plasma sheet, see>{Artemyev22:jgr:ELFIN&THEMIS}, the relativistic electron burst measured by CIRBE was not embedded into the much more extended plasma sheet, unlike ELFIN observations. Therefore, in CIRBE data, the relativistic electron bursts in the magnetotail are not as vibrant compared to the outer radiation belt fluxes, and we mark them with red arrows. 

The event in Fig. \ref{fig3}(a) was in conjunction with POES NOAA-19 observations: Fig.  \ref{fig3}(b) shows that the relativistic burst occurred well poleward of the electron isotropy boundary (at $\sim62.5^\circ$ magnetic latitude), as implied from the precipitating and trapped electron ($>30$keV) fluxes at POES \cite<we use AACGM magnetic latitudes for both CIRBE and POES, see>{AACGM}. This relativistic electron burst was observed when THEMIS E \cite{Angelopoulos08:ssr} was in the night-side magnetosphere and detected enhanced energetic electron fluxes. Panels (d-f) show magnetic field perturbations \cite{Auster08:THEMIS}, plasma flow speed \cite{McFadden08:THEMIS}, and energetic electron fluxes \cite{Angelopoulos08:sst}. The interval delineated by the vertical dashed lines correspond to near-equatorial measurements (implied by the $|B_x|$ decrease) following a plasma injection around 15:30 UT (better seen in THEMIS A and D; not shown). This injection results in a magnetotail dipolarization ($|B_x|$ decreases, $B_z$ increases) and energetic flux enhancement. Therefore, CIRBE observations are associated with substorm activity (panel (c) shows the AE index increase around 17:00 UT), similar to ELFIN observations shown in Figs. \ref{fig1}, \ref{fig2}. Also, similar to the ELFIN/MMS comparison, THEMIS does not detect an increase in relativistic electron fluxes (the main flux increase is below $500$keV). However, a comparison of $<500$keV energy spectra measured by CIRBE and THEMIS shows a strong similarity (see Fig. \ref{fig3}(g)), confirming CIRBE's projection to the post-injection magnetotail, close to the location of THEMIS E. Despite several independent confirmations that the CIRBE measurements at MLAT$<-64^\circ$ originate from the plasma sheet--such as their location relative to the isotropy boundary observed by POES and spectrum validation with THEMIS--we cannot definitively pinpoint the location of the relativistic bursts detected by CIRBE without direct measurements of precipitating flux. These bursts could potentially still represent the poleward edge of the outer radiation belt. Therefore, systematic analysis of more events of this kind is necessary to better understand their true nature.

\begin{figure}[!htbp]
    \centering
    \includegraphics[width=1.0\linewidth]{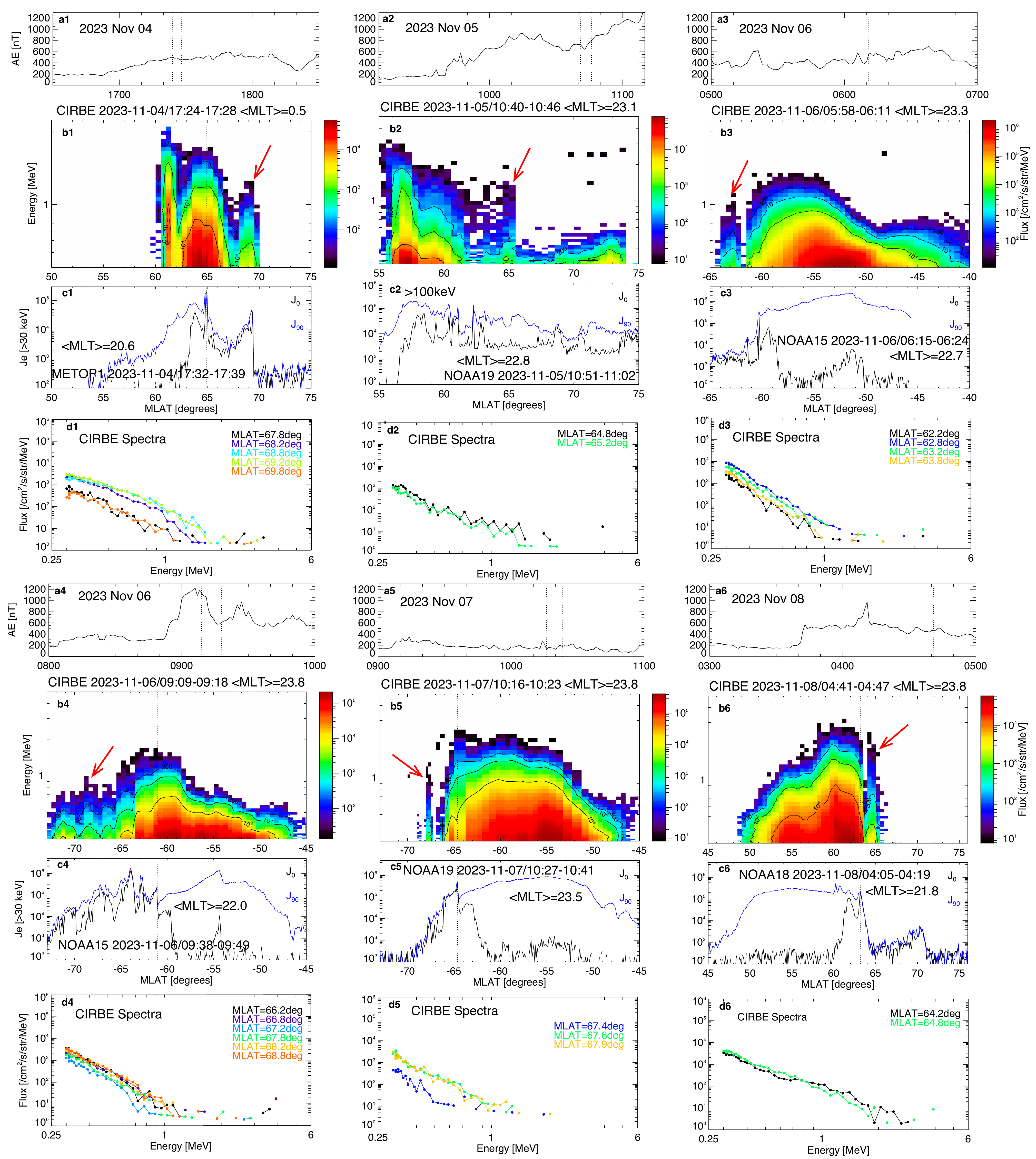}
    \caption{Overview of six CIRBE events with relativistic electron observations in near-Earth magnetotail: 4 November 2023, at 17:30UT; 4 November 2023, at 19:30UT; 5 November 202, at 10:40UT, 6 November 2023, at 06:00UT; 6 November 2023, at 09:20UT; and 7 November 2023, at 10:20UT. For each event, we show (a) AE index during the 2h interval embedding CIRBE observations, (b) CIRBE measurements of locally trapped fluxes, (c) POES measurements of trapped and precipitating fluxes of $>30$keV electrons, and (d) the 1D electron spectra measured by CIRBE during relativistic electron bursts in the magnetotail.}
    \label{fig4}
\end{figure}

The energy spectrum of the relativistic electron burst in Fig. \ref{fig3}(g) exhibits a power-law shape, in contrast to a clear exponential shape within the outer radiation belt. Fitting to the function that combines power-law and exponential components:  $A\cdot E^{-\alpha}\cdot\exp(-E/E_0)$, we derive values for the mean energy $E_0$ and power-law index $\alpha$. For the magnetotail, we find $E_0\approx 400$keV (i.e., the exponential factor $\sim \exp(-E/E0)$ exerts a negligible influence on the spectrum at $<2$MeV) and $\alpha\approx 4$; on the other hand, for the outer radiation belt, we have $E_0\approx 130$keV (indicating a pronounced exponential decay at high energies), $\alpha\approx 0$. Thus, there is a distinct difference between the exponential spectrum in the outer radiation belt (characterized by the mean energy $E_0$ of the exponential decay, $\sim \exp(-E/E_0)$) and the power-law spectrum in the magnetotail relativistic burst (dictated by the index $\alpha$). This power-law spectrum is typical for such relativistic populations: Fig. \ref{fig4} shows six more examples of substorm-associated relativistic electron bursts with such power-law spectra. In all events, the relativistic bursts are observed poleward of the outer radiation belt (the boundary between the outer radiation belt and the magnetotail can be discerned from measurements at a nearby POES satellite, as seen in the precipitating and trapped fluxes of $>30$keV electrons), with energies exceeding $1$MeV. The typical power-law index $\alpha$ for these events varies within the $[1.5,4.5]$ range, with $E_0>500$keV making the factor $\sim \exp(-E/E_0)$ insignificant for $<2$MeV energies. The power-law spectrum with the exponential cut-off is expected for electron acceleration in a single reconnection region \cite<see details in>{Bulanov76,Burkhart90,Birn12:SSR}, whereas multiple (turbulent) reconnection would produce a power-law energy tail \cite{Hoshino12,Johnson22:apj_rx,Oka22:rx}. Thus, we suggest that the observed relativistic electron populations were initially accelerated in near-Earth turbulent reconnection region \cite<distributions observed around the near-Earth reconnection region usually show upper energies at $\sim 100$keV, see>{Maha11NatPh,Zhou11:lsk,Birn13:jgr,Usanova&Ergun22}, and then adiabatically or quasi-adiabatically heated during Earthward injection. Conservation of the first adiabatic moment has been shown to provide a factor of $\times 10$ in energy for near-Earth injections \cite<see>{Birn14,Sorathia18,Eshetu19} and will likely play an important role for any Earthward motion. However, since the observed electrons precipitate tailward from the injection region, conservation of the second adiabatic moment as field lines contract may also play an important role. It is well known that field lines in contracting flux ropes can energize particles through Fermi acceleration and create power-laws \cite<see>{Arnold21,Li21:pop,Zhang21:prl,Nakanotani22}. In the magnetotail a similar process is likely at play from particles mirroring off the near-Earth magnetic field on flux ropes associated with fast flows that are contracting and moving Earthward. This combination may potentially explain the observed relativistic electron bursts at $\sim 1$MeV \cite<see also discussion in>{Turner21:grl_mms&rbsp}, which still requires detailed quantification, especially for the more extreme events reaching $\sim 5$MeV in Fig. \ref{fig2}.

\section{Conclusions}  
\label{sec:conclusions}
In this study, we have analyzed relativistic and super-relativistic electron events in near-Earth magnetotail captured by low-altitude CubeSats. These bursts of relativistic electrons exhibit the following properties:
\begin{itemize}
  \item Electron energies within these bursts can reach the relativistic ($\sim 1$MeV) and even ultra-relativistic ($\sim 3$MeV) range
  \item These bursts are well separated from the outer radiation belt and are thus likely related to acceleration mechanisms in the magnetotail.
  \item These are spatially and temporally localized bursts.
  \item Observed poleward from the electron isotropy boundary, these bursts are likely associated with an isotropic (highly scattered) electron population, indicative of relativistic electron precipitation into the polar cap.
\end{itemize}

Earth's magnetotail is a spatially compact system dominated by thermal electron energies below $\sim1$keV \cite<e.g.,>{Artemyev12:jgr:electrons}. Effective acceleration mechanisms are thus required to explain the observed relativistic electron population. Magnetic field line reconnection in the magnetotail current sheet alone may not produce these electrons \cite<see theoretical estimates>{Bulanov76,Burkhart90,Birn12:SSR,Egedal12}, but it can generate the {\it seed} electron population at $\sim 10-100$keV \cite<see>[and references therein]{Turner21:reconnection}, which can be further adiabatically accelerated during Earthward transport \cite{Maha11NatPh,Maha13:lsk,Pan12}. Although electrons of such energies are not magnetized in the magnetotail and do not preserve their adiabatic invariants, these demagnetized electrons can still experience adiabatic acceleration with energy scaled with the equatorial magnetic field intensity, $\propto B_{eq}^\kappa$ \cite<the power-law exponent, $\kappa<1$, depends on the specific magnetic field configuration, see details in>[and references therein]{Zelenyi13:UFN}. Therefore, the presented observations \cite<along with historical datasets collected by ISEE-3 and Imp 8; see>{Baker&Stone77,Slavin92,Richardson93}  demonstrate the potential for producing relativistic ($\geq 1$MeV) electrons within the plasma systems of spatial scales $\leq 10^2 \rho_e$: the diameter of Earth's magnetotail is $\sim 40 R_E$, which is smaller than 100 times the electron gyroradii ($\rho_e$) at $\geq 1$MeV and $\sim 1$nT equatorial field. This makes the magnetotail (and the magnetic reconnection region) an extremely efficient yet compact accelerator. These observations, along with the future developed acceleration models and resolving the question of plasma sheet ($\sim 1$keV) electron pre-acceleration up to $50-100$ keV before undergoing adiabatic acceleration to $1$MeV, may be very important, because magnetic reconnection is responsible for the transformation of magnetic field energy into plasma heating and charged particle acceleration in a wide range of space plasma systems \cite<e.g.,>[and references therein]{Zweibel&Yamada09,Hoshino&Lyubarsky12,Ji22:NatRP,Guo24:ssr}.

\acknowledgments
We acknowledge the CIRBE mission team for their efforts in making the dataset accessible to the community, under NASA Heliophysics Division Grants 80NSSC19K0995 and 80NSSC21K0583. We are grateful to NASA's CubeSat Launch Initiative for ELFIN's successful launch in the desired orbits. We acknowledge early support of ELFIN project by the AFOSR, under its University Nanosat Program, UNP-8 project, contract FA9453-12-D-0285, and by the California Space Grant program. We acknowledge critical contributions by numerous ELFIN team members and support by NASA 80NSSC22K1005 and NSF grants AGS-1242918, AGS-2019950. We acknowledge NASA contract NAS5-02099 for the use of data from the THEMIS Mission. We thank the entire MMS team and instrument leads for data access and support. 

A.V.A and X.-J. Z. acknowledge support from the NASA grants 80NSSC23K0108, 80NSSC23K0413, 80NSSC24K0561. X.-J. Z. acknowledges support from the NASA grant 80NSSC24K0138 and the NSF grant 2329897. V.A. acknowledges support by NSF award AGS-2019950 and NASA contract NAS5-02099. M.S. was supported by the MMS via subcontract to the SwRI (NNG04EB99C).

\section*{Open Research} \noindent 
Fluxes measured by ELFIN are available in ELFIN data archive https://data.elfin.ucla.edu/ in CDF format.\\ \noindent
Fluxes measured by CIRBE are available in CIRBE data archive https://lasp.colorado.edu/cirbe/data-products/ in NCDF format.\\ \noindent
THEMIS and MMS data are available at \url{http://themis.ssl.berkeley.edu} and \url{https://lasp.colorado.edu/mms}.

Data analysis was done using SPEDAS V4.1 \cite{Angelopoulos19}. The software can be downloaded from {http://spedas.org/wiki/index.php?title=Downloads\_and\_Installation}.



\end{document}